\documentstyle[12pt]{article}
\catcode`\@=11
\long\def\@makefntext#1{
\protect\noindent \hbox to 3.2pt {\hskip-.9pt
$^{{\ninerm\@thefnmark}}$\hfil}#1\hfill}                

 \def\@makefnmark{\hbox to 0pt{$^{\@thefnmark}$\hss}}  

\def\ps@myheadings{\let\@mkboth\@gobbletwo
\def\@oddhead{\hbox{}
\rightmark\hfil\ninerm\thepage}
\def\@oddfoot{}\def\@evenhead{\ninerm\thepage\hfil
\leftmark\hbox{}}\def\@evenfoot{}
\def\sectionmark##1{}\def\subsectionmark##1{}}

\newcounter{sectionc}\newcounter{subsectionc}\newcounter{subsubsectionc}
\renewcommand{\section}[1] {\vspace{0.6cm}\addtocounter{sectionc}{1}
\setcounter{subsectionc}{0}\setcounter{subsubsectionc}{0}\noindent
	{\bf\thesectionc. #1}\par\vspace{0.4cm}}
\renewcommand{\subsection}[1] {\vspace{0.6cm}\addtocounter{subsectionc}{1}
	\setcounter{subsubsectionc}{0}\noindent
	{\it\thesectionc.\thesubsectionc. #1}\par\vspace{0.4cm}}
\renewcommand{\subsubsection}[1]
{\vspace{0.6cm}\addtocounter{subsubsectionc}{1}
	\noindent {\rm\thesectionc.\thesubsectionc.\thesubsubsectionc.
	#1}\par\vspace{0.4cm}}

\newcounter{appendixc}
\newcounter{subappendixc}[appendixc]
\newcounter{subsubappendixc}[subappendixc]

\renewcommand{\appendix}[1] {\vspace{0.6cm}
	\refstepcounter{appendixc}
	\setcounter{figure}{0}
	\setcounter{table}{0}
	\setcounter{equation}{0}
	\renewcommand{\thefigure}{\Alph{appendixc}.\arabic{figure}}
	\renewcommand{\thetable}{\Alph{appendixc}.\arabic{table}}
	\renewcommand{\theappendixc}{\Alph{appendixc}}
	\renewcommand{\theequation}{\Alph{appendixc}.\arabic{equation}}
	\noindent{\bf Appendix \theappendixc #1}\par\vspace{0.4cm}}

\def\abstracts#1{{
	\centering{\begin{minipage}{30pc}\footnotesize\baselineskip=12pt\noindent
	\centerline{\tenrm ABSTRACT}\vspace{0.3cm}
	\parindent=0pt #1
	\end{minipage}}\par}}

\renewenvironment{thebibliography}[1]
	{\begin{list}{\arabic{enumi}.}
	{\usecounter{enumi}\setlength{\parsep}{0pt}
\setlength{\leftmargin 1.25cm}{\rightmargin 0pt}
	 \setlength{\itemsep}{0pt} \settowidth
	{\labelwidth}{#1.}\sloppy}}{\end{list}}

\newcounter{itemlistc}
\newcounter{romanlistc}
\newcounter{alphlistc}
\newcounter{arabiclistc}

\newcommand{\fcaption}[1]{
	\refstepcounter{figure}
	\setbox\@tempboxa = \hbox{\footnotesize Fig.~\thefigure. #1}
	\ifdim \wd\@tempboxa > 6in
	   {\begin{center}
	\parbox{6in}{\footnotesize\baselineskip=12pt Fig.~\thefigure. #1}
	    \end{center}}
	\else
	     {\begin{center}
	     {\footnotesize Fig.~\thefigure. #1}
	      \end{center}}
	\fi}

\newcommand{\tcaption}[1]{
	\refstepcounter{table}
	\setbox\@tempboxa = \hbox{\tenrm Table~\thetable. #1}
	\ifdim \wd\@tempboxa > 6in
	   {\begin{center}
	\parbox{6in}{\tenrm\baselineskip=12pt Table~\thetable. #1}
	    \end{center}}
	\else
	     {\begin{center}
	     {\tenrm Table~\thetable. #1}
	      \end{center}}
	\fi}

\def\@citex[#1]#2{\if@filesw\immediate\write\@auxout
	{\string\citation{#2}}\fi
\def\@citea{}\@cite{\@for\@citeb:=#2\do
	{\@citea\def\@citea{,}\@ifundefined
	{b@\@citeb}{{\bf ?}\@warning
	{Citation `\@citeb' on page \thepage \space undefined}}
	{\csname b@\@citeb\endcsname}}}{#1}}

\newif\if@cghi
\def\cite{\@cghitrue\@ifnextchar [{\@tempswatrue
	\@citex}{\@tempswafalse\@citex[]}}
\def\citelow{\@cghifalse\@ifnextchar [{\@tempswatrue
	\@citex}{\@tempswafalse\@citex[]}}
\def\@cite#1#2{{$\null^{#1}$\if@tempswa\typeout
	{IJCGA warning: optional citation argument
	ignored: `#2'} \fi}}

\def\fnt#1#2{\footnotetext{\kern-.3em
	{$^{\mbox{\sevenrm #1}}$}{#2}}}

 1
 1
 1

\font\tenbf=cmbx10
\font\tenrm=cmr10
\font\tenit=cmti10

\font\ninerm=cmr9

\def\aleq{\vcenter{\vbox{\hbox{$\buildrel < \over \sim$}}}}
\def\ageq{\vcenter{\vbox{\hbox{$\buildrel > \over \sim$}}}}
\newcommand{\be}{\begin{equation}}
\newcommand{\ee}{\end{equation}}
\newcommand{\bes}{\begin{eqnarray}}
\newcommand{\ees}{\end{eqnarray}}
\newcommand{\nummer}[1]{\hfill #1 \par}

\newcommand{\plotfolder}{/tmp_mnt/hosts/cleopatra/u1/people/wos}
\newcommand{\FIG}[4]{
\hfil \hbox to #2{\vbox to #3{\vfil
                        \special{ps: epsfile \plotfolder/#1 #4}
} }
\hfil
}

\begin{document}
\nummer{UNIGRAZ-UTP}
\nummer{25-01-95}
\vspace{1.5 truecm}
\centerline{\tenbf EXCLUSIVE PHOTON-INDUCED HADRONIC REACTIONS}
\baselineskip=22pt
\centerline{\tenbf AT LARGE MOMENTUM TRANSFERS
\footnote{Talk given at the 7th Adriatic Meeting on Particle Physics, Brijuni
(Croatia),
September 1994}}
\baselineskip=16pt
\vspace{0.65cm}
\centerline{\tenrm W.SCHWEIGER
\footnote{E-mail: schweigerw@edvz.kfunigraz.ac.at}}
\baselineskip=13pt
\centerline{\tenit Institute of Theoretical Physics, University of Graz,
Universit\"atsplatz 5}
\baselineskip=12pt
\centerline{\tenit A-8010 Graz, Austria}
\vspace{1.3truecm}
\abstracts{
It is generally assumed that due to factorization of long- and short-distance
dynamics
perturbative QCD can be applied to exclusive hadronic reactions at large
momentum
transfers. Within such a perturbative approach diquarks turn out to be a useful
phenomenological device to model non-perturbative effects still observable in
the  kinematic
range accessible by present-days experiments. The basic ingredients of the
perturbative
formalism with diquarks, i.e. Feynman rules for diquarks and quark-diquark wave
functions of baryons, are briefly summarized. Applications of the diquark model
to the
electromagnetic form factors of the proton in the space- as well as time-like
region,
Compton-scattering off protons, $\gamma \gamma
\longrightarrow p \bar{p}$, and photoproduction of Kaons are discussed.}

\vspace{1.0truecm}
\rm\baselineskip=14pt
\section{Introduction and Motivation}
\vspace{-0.25cm}
Despite the big progress achieved in strong-interaction physics since the
advent of QCD, in
particular in the understanding of deep inelastic scattering~\cite{Sti94}, the
appropriate
theoretical description of {\bf exclusive hadronic reactions} for momentum
transfers
reaching from a few GeV up to the highest values accessible by present-days
experiments
remains still a challenging problem. This is just the kinematic range where the
transition from the non-perturbative to the perturbative regime of QCD is
commonly believed to
take place. Therefore the physical picture of such reactions is not as
clear-cut as one may
hope. The situation is reflected by the fact that some of the experimentally
measured exclusive
observables exhibit features typical for perturbative QCD already at a few GeV
of momentum
transfer, whereas others do not.  Features characteristic for pQCD are in
particular
{\bf suppression of hadronic helicity flips}~\cite{Bro81}
and
(fixed angle) {\bf power laws}~\cite{Bro75a} obeyed by exclusive scattering
amplitudes
at large momentum transfer $Q$
 -- in electron-proton scattering, e.g., the Dirac form factor $F_1^p$ is
expected to
scale like $Q^{-4}$.
Both these features are a consequence of the basic assumption entering the pQCD
approach, namely {\bf factorization of long- and short-distance
dynamics}~\cite{Le80a,Ef80a}.

According to the theoretical framework emerging from the factorization
hypothesis, the so called ``Hard Scattering Picture'' (HSP),  an exclusive
hadronic amplitude
$T$ can be expressed as a convolution of process independent distribution
amplitudes (DAs)
$\phi_{H_i}$ with a perturbative hard amplitude
$\widehat{T}$ for the scattering of collinear constituents
\be
T = \int_0^1 \widehat{T}(x_j,Q) \, \prod_{H_i} \, \left(
\phi_{H_i}(x_j,\tilde{Q})
            \, \delta(1 - \sum_{k=1}^{n_i} x_k) \, \prod_{j=1}^{n_i} dx_j
\right) \; .
\ee
To leading order in $(1/Q)$ only tree-graphs, obtained by replacing each of the
external
hadrons by its valence Fock state, contribute to $\widehat{T}$. The
distribution amplitudes
$\phi_{H_i}$ are probability amplitudes for finding the pertinent valence
Fock-state in the
hadron $H_i$ with the constituents carrying the longitudinal momentum fractions
$x_j$ of the
parent hadron and being collinear up to the (factorization) scale $\tilde{Q}$.

As a simple illustration for the apparent successes and also the limitations of
the HSP let
me mention (once more) the electromagnetic form factors of the proton:
\begin{itemize}
\item on the one hand, the data for the helicity-non-flip Dirac form factor
$F_1^p$
indeed seem to scale according to perturbative QCD for $Q^2 \ageq 10
\hbox{GeV}^2$~\cite{Si93a}
(cf. Fig.1);
\item on the other hand, the helicity-flip form factor $F_2^p$ is still sizable
around
$10\, \hbox{GeV}^2$,
$
\left. ({Q^2 F_2^p(Q^2)}/{F_1^p(Q^2)}) \right\vert_{Q^2 = 8.83
\hbox{\scriptsize GeV}^2} =
1.143^{+0.821}_{-0.403} \hbox{GeV}^2
$~\cite{And94a},
violating helicity conservation. The pure quark HSP only states that $F_2^p$
should be
suppressed as compared to $F_1^p$ by a factor $\tilde{m}^2/Q^2$ (where
$\tilde{m}$ is some soft
scale) -- the recent data on $F_1^p$ and $F_2^p$ indeed seem to confirm such a
behaviour (cf.
Fig.~2). However, the pure quark HSP cannot provide any quantitative
predictions for $F_2^p$.
\end{itemize}
\noindent
Also in other exclusive hadronic reactions, like elastic p-p
scattering~\cite{Cra90a}, violation
of hadronic helicity conservation is observed which amounts up to $\approx
20-30\%$ even at a
few GeV of momentum transfer. Among others these findings can be regarded as a
hint that
soft  contributions not accounted for by the pure quark HSP are still present
in the few-GeV
region.

In order to cope with such effects a phenomenological model has been proposed
in a
series of papers~\cite{Kro91a,Kro91b,Ja93a,Kro93a} which still pursues the
perturbative
approach, in which baryons, however, are treated as quark-diquark systems. This
picture is
primarily motivated by the observation that most of the octet-baryon DAs
derived via QCD
sum-rules (at finite $\tilde{Q}$) are very asymmetric~\cite{Che89a},
indicating strong quark-quark correlations.  Actually, a very asymmetric DA,
which strongly
favours the helicity-parallel u quark, is needed for the proton to achieve a
quantitative
description of $F_1^p$ for $Q^2 \ageq 10$GeV$^2$~\cite{Ste89}. For the
completely symmetric DA
$\propto x_1 x_2 x_3$, which emerges in the limit $\tilde{Q} \rightarrow
\infty$, the form
factor $F_1^p$ becomes identically zero.

\section{The Hard-Scattering Picture with Diquarks}
As has been mentioned already, there are two elements entering the calculation
of an
exclusive amplitude within the HSP, namely a hard scattering amplitude
$\widehat{T}$, to be
calculated in collinear approximation within perturbative QCD, and DAs $\phi$.
To leading order
in $1/Q$ the scattering amplitude is determined by the valence Fock states of
the hadrons
under consideration -- in the diquark model a {\bf quark-diquark} state in case
of an ordinary
baryon. If one assumes the diquark to be in its ground state it can be either a
spin 0 (scalar)
diquark or a spin 1 (axial vector) diquark. The introduction of vector diquarks
occurs to be
essential if one wants to describe spin effects. By the way, there are many
fields of hadronic
physics, like baryon spectroscopy, deep inelastic scattering, or weak decays,
to mention a
few, in which the concept of diquarks has already been applied successfully.
Recently, a
comprehensive review of the main ideas about diquarks has been published by
M.Anselmino et
al.~\cite{Ans93a}.

To come back to exclusive scattering, let us continue with baryon wave
functions in terms
of the quark and diquark constituents. For the lowest lying baryon octet,
assuming zero
relative orbital angular momentum between quark and diquark the wave functions
(already
integrated over transverse momenta) can be cast into the form
\be
\vert B; p, \lambda \rangle =  f_S \phi_S^B (x_1) \,
\chi^B_S u( p,\lambda ) +
   {(f_V / \sqrt{3})}  \phi_V^B (x_1) \, \chi^B_V \, (\gamma^\alpha + p^\alpha
/ m_B )
    \, \gamma_5 u( p,\lambda )  \, . \label{wfcov}
\ee
The two terms in Eq.(\ref{wfcov}) represent configurations consisting of a
quark and
either a scalar (S) or vector (V) diquark. Thereby advantage has been taken of
the collinear
situation, $p_q = x_1 p$ and $p_D = x_2 p = (1 - x_1) p$ ($x_1$ is the
longitudinal momentum
fraction  of the quark) to rewrite the spin part of the wave function in a
covariant
way. The pleasant feature of the wave-function representation Eq.(\ref{wfcov})
is that it
contains, besides $x_1$, only baryonic quantities (momentum $p$, helicity
$\lambda$, baryon
mass $m_B$).

For an SU(6)-like spin-flavour dependence, which means starting from the SU(6)
wave function
for three quarks and combining, let us say, quark 2 and 3 to a diquark, the
flavour functions
$\chi$ for proton and, e.g., $\Lambda$ take on the form
\be
\chi_S^p = u S_{[u,d]} \, , \;\;\;
\chi_V^p = \phantom{-} [u V_{\{u,d\}} -\sqrt{2} d V_{\{u,u\}}] / \sqrt{3} \, ,
\ee
\be
\chi_S^{\Lambda^0} = [ u S_{[d,s]} - d S_{[u,s]} - 2 s S_{[u,d]}] / \sqrt{6}
\, ,
\;\;\;
\chi_V^{\Lambda^0} =  [u V_{\{d,s\}}- d V_{\{u,s\}}] / \sqrt{2} \, .
\ee

The DAs $\phi_S^B(x_1)$ and $\phi_V^B(x_1)$ in Eq.(\ref{wfcov}) are nothing
else but
quark-diquark light-cone wave functions integrated over the transverse
momentum. $f_S$ and
$f_V$ are the $r=0$ values of the corresponding configuration space wave
functions. Since the
model is applied to a very restricted momentum-transfer range ($3 - 30$
GeV$^2$) the QCD
evolution of these DAs, which brings in only a weak logarithmic $\tilde{Q}$
dependence, is
neglected. In general, very little is known on the shape of quark-diquark DAs.
Hence one is
forced to make a phenomenological ansatz for which an expression of the form
\be
\phi_D^B (x_1) = N_D x_1 x_2^3 (1 + c_{1 D} x_1 + c_{2 D} x_1^2)
\exp \left[ - b^2 \left( {m_q^2 / x_1} + {m_D^2 / x_2} \right) \right] \; ,
\label{DAp}
\ee
where $D$ can be either $S$ or $V$, proves to be quite appropriate. Actually,
the constants
$c_{1 S}$ and $c_{2 S}$ are fixed to zero. For $c_{1 D} = c_{2 D} = 0$ the form
of the DA
(\ref{DAp}) can be traced back to a nonrelativistic harmonic-oscillator wave
function~\cite{Hua89}. Therefore the masses appearing in the exponentials have
to be considered
as constituent masses (330 MeV for light quarks, 580 MeV for light diquarks,
strange quarks
are 150 MeV heavier than light quarks). The oscillator parameter $b^2 = 0.248
\hbox{GeV}^{-2}$
is chosen in such a way that the full wave function gives rise to a value of
$600$ MeV for the
mean intrinsic transverse momentum of quarks inside a nucleon. The
``normalization constants''
$N_S$ and $N_V$ are determined by the condition $
\int_0^1 dx_1 \, \phi_{S(V)}(x_1) = 1 \; .
$

Concerning colour, the diquark behaves like an antiquark. In order to give a
colourless baryon in combination with a colour triplet quark the diquark has to
be in
a colour antitriplet state.

The ingredients of the perturbative part of the model are gluon-diquark and
photon-diquark
vertices which are introduced following the standard prescriptions for the
coupling of a spin-1 gauge boson to a spin-0 or a spin-1 particle,
respectively~\cite{Ja93a}.
Thereby V diquarks are allowed to posses an anomalous (chromo)magnetic moment
$\kappa_V$.
In applications of the model Feynman graphs are calculated first with these
Feynman rules for
point-like diquarks. In order to take into account the composite nature of
diquarks the n-point
contributions, i.e. those Feynman graphs where $n-2$ gauge bosons couple to the
diquark, are
multiplied afterwards with phenomenological vertex functions -- diquark form
factors. The
particular choice
\be F_S^{(3)} (Q^2) = \; \delta_S \, {Q_S^2 / (Q_S^2 +
Q^2)} \; ,  \qquad F_V^{(3)} (Q^2) = \; \delta_V \left( {Q_V^2 / (Q_V^2 +
Q^2)}\right)^2
\ee  %
for 3-point functions and
\be
F_S^{(n)} = a_S F_S^{(3)} (Q^2) \, , \;\;\;\;
F_V^{(n)} = a_V F_V^{(3)} (Q^2) \left( {Q_V^2 /( Q_V^2 + Q^2 )}\right)^{(n-3)}
\ee
for n-point functions ($n \geq 4$) ensures that in the limit $Q^2 \rightarrow
\infty$ the scaling behaviour of the diquark model goes over in that of the
pure
quark HSP. If one knew the quark-quark distribution amplitude of a diquark one
could
in principle calculate these form factors from the pure quark HSP.
The factor $\delta_{S (V)}  =  \alpha_s (Q^2) / \alpha_s (Q^2_{S (V)})$
($\delta_{S (V)} = 1$ for $Q^2 \leq  Q^2_{S (V)}$) provides the correct powers
of
$\alpha_s (Q^2)$ for asymptotically large $Q^2$. For the running coupling
$\alpha_S$ the
one-loop result
$\alpha_S = 12 \pi / 25 \ln (Q^2 / \Lambda_{QCD}^2 )$
is used with $\Lambda_{QCD} = 200 {\rm \hbox{MeV}}$. In addition $\alpha_S$ is
restricted
to be smaller than $0.5$. $a_S$ and $a_V$ are strength parameters which allow
for the
possibility of diquark excitation and break-up in intermediate states where
diquarks can
be far off-shell.

\section{Parameter Fixing and Electromagnetic Form Factors of the
Proton~\cite{Ja93a}}
The open parameters of the model are the cut-off masses $Q_S^2$ and $Q_V^2$
occurring in the
diquark form factors, the anomalous (chromo)magnetic moment of the vector
diquark
$\kappa_V$, the strength parameters of the n-point functions $a_S$ and $a_V$,
and the
parameters in the diquark DAs $f_S$, $f_V$, $c_{1V}$, and $c_{2V}$.
These parameters have been determined by means of e-p and e-n scattering data
in the range
3 GeV$^2$ $\aleq$ Q$^2$ $\aleq$ 30 GeV$^2$, namely
$(d\sigma/d\Omega)_{ep}$~\cite{Si93a},
$G_E^p$ and $G_M^p$~\cite{And94a} (extracted via Rosenbluth separation),
$\sigma_n/\sigma_p$~\cite{Ro92},
$G_E^n$ and $G_M^n$~\cite{Lu93}.

A good fit to the data is accomplished with the parameter set
\bes
f_S = 73.85 \hbox{MeV}, \; Q_S^2 = 3.22 \hbox{GeV}^2, \; a_S = 0.15, &   \\
f_V = 127.7 \hbox{MeV}, \; Q_V^2 = 1.50 \hbox{GeV}^2, \; a_V = 0.05, &
\; \kappa_V = 1.39, \; c_1 = 5.8, \; c_2 = -12.5 .  \nonumber
\ees
The results for the proton magnetic form factor $G_M^p$ and the Pauli form
factor $F_2^p$
are depicted in Figs.~1 and 2, respectively. It should be emphasized that
unlike the pure quark HSP, which cannot make any quantitative predictions for
the
helicity-flip Pauli form factor $F_2^p$, the diquark model provides reasonable
results also
for this quantity.

\begin{figure}
\vspace{7.5 truecm}
\fcaption{The magnetic form factor of the proton in the space-like and
time-like
$(s = -Q^2)$
regions. The solid lines represent the corresponding predictions of the diquark
model. The
space-like data are taken from Sill et al.~\protect{\cite{Si93a}}($\bullet$),
the time-like data are
taken from Bardin et al.~\protect{\cite{Bar94}}($+$), Antonelli et
al.~\protect{\cite{Ant94}}
($\Box$), and
Armstrong et al.~\protect{\cite{Arm93}}($\circ$).
\label{gmp}}
\vspace{-0.5 truecm}
\end{figure}

\begin{figure}
\vspace{7.5 truecm}
\fcaption{The Pauli form factor of the proton $F^2_p$ scaled by $Q^6$. The
solid line
represents the result obtained with the diquark model. Data are taken from
Andivalis et
al.~\protect{\cite{And94a}} ($\bullet$).
\label{f2p}}
\end{figure}

After having fixed the parameters of the diquark model by means of elastic e-N
scattering we
are now in the position to see how it works for other exclusive reactions.

\section{Applications of the Diquark Model}
\vspace{-0.7cm}
\subsection{Proton Magnetic Form Factor in the Timelike Region~\cite{Kro93a}}
\vspace{-0.35cm}
The nearest application at hand is the continuation of the nucleon form factors
to the
timelike region, i.e. to consider in particular e$^+$ e$^-$ $\longrightarrow$ p
$\bar{\hbox{p}}$   or p $\bar{\hbox{p}}$ $\longrightarrow$ e$^+$ e$^-$
scattering. These
reactions are already a serious challenge for the pure quark HSP, which
predicts
$ G_M^p (s) = G_M^p (Q^2 = -s)$.
By way of contrast, recent measurements performed by the E760 collaboration at
Fermilab~\cite{Arm93} get results for the proton magnetic form factor in the
timelike region which
are more than 2 times larger than the spacelike values, even for $Q^2 \ageq 10$
GeV$^2$.
Thus it seems very hard for the pure quark HSP to reconcile the space- and
timelike
predictions. Some improvement has been achieved by Hyer~\cite{Hy93}, who took
into account
radiative (Sudakov) corrections, but a more or less sizable discrepancy --
depending on the
proton DA employed -- is still observable.

Applying the diquark model to timelike processes requires some care with
respect
to the diquark form factors. One has to keep in mind that the multipole
parameterization of the diquark form factors is only an effective description
of the
composite nature of diquarks. It is not possible to continue this
parameterization in a unique
way to the timelike region since the exact dynamics of the diquark system is
not known. The
simple analytic continuation $Q^2 \longrightarrow -s$ indeed leads to poles in
the diquark
form factors which have no physical meaning. In order to avoid such unphysical
poles the
diquark form factors are kept constant once they reach a certain value, say
$c_0$, i.e.
\be
F_D^{(n)} (s) =  F_D^{(n)} (-Q^2)\, ,  \quad  s \geq s_0 \quad \hbox{and} \quad
F_D^{(n)} (s) = c_0 = F_D^{(n)} (s_0)\, ,  \quad  s < s_0J \, . \label{formt}
\ee
A similar recipe has also been used in a study of (inclusive) e$^+$-e$^-$
annihilation
into hadrons~\cite{Ek84}.
{}From Eq.(\ref{formt}) it is evident that the diquark form factors and hence
also the proton
magnetic form factor have larger values for a given $s$ ($>0$) than for the
corresponding
$Q^2 = -s$. In the timelike region one is closer to the form factor singularity
than in
the spacelike domain. This resembles the situation in the vector-meson
dominance model.

$c_0$ is of course a new parameter of the model. But once it is fixed the
diquark
model should also be applicable to other timelike reactions. One finds good
agreement with the
E-760 data on $G_M^p$ for $c_0 = 1.3$ (cf. Fig.1). As will be seen in Sect.4.3
this
prescription also gives reasonable results for $\gamma \gamma \longrightarrow
\hbox{p}
\bar{\hbox{p}}$.
%

\subsection{Compton Scattering off Protons~\cite{Kro91b}}
\vspace{-0.35cm}
For Compton scattering  differential
cross section data are available up to  $\vert t \vert \approx 4 GeV^2$.
$\gamma \hbox{p} \rightarrow \gamma \hbox{p}$ is usually described by 6
independent (cms) helicity amplitudes
\be
\begin{array}{llllll}
\phi_1 &= M_{1 {1 \over 2},1 {1 \over 2}} & \sim  s^{-2} \; , &\hspace{1.0cm}
           \phi_2 &= M_{-1 -{1 \over 2},1 {1 \over 2}} & \sim  s^{-5/2} \; ,\\
\phi_3 &= M_{-1 {1 \over 2},1 {1 \over 2}} & \sim  s^{-3} \; , &\hspace{1.0cm}
           \phi_4 &=  M_{1 -{1 \over 2}, 1 {1 \over 2}} & \sim  s^{-5/2} \; ,\\
\phi_5 &=  M_{1 -{1 \over 2}, 1 -{1 \over 2}} & \sim s^{-2} \; ,
&\hspace{1.0cm}
           \phi_6 &=  M_{-1 {1 \over 2}, 1 -{1 \over 2}} & \sim s^{-5/2} \; .
\\
\end{array}\label{comamp}
\ee
In Eq.(\ref{comamp}) the (fixed angle) large s behaviour of the helicity
amplitudes, as
resulting from the diquark model, is already indicated. $\phi_1$, $\phi_3$ and
$\phi_5$
conserve the helicity of the proton, whereas the others do not.

\begin{figure}
\vspace{10 truecm}
\vspace{-1.0 truecm}
\fcaption{The Compton cross section vs. $\cos(\vartheta)$ for various photon
lab. energies.
Data are taken from Shupe et al.~\protect{\cite{Shu79}}. Solid line:
diquark-model result
at 4 GeV ($s = 8.4$ GeV$^2$); dashed line: prediction of the pure quark
HSP~\protect{\cite{Kron91}}.
\label{compton}}
\end{figure}

A new interesting  feature showing up in Compton scattering is the occurrence
of a
relative phase between helicity flip and non-flip amplitudes. This comes about
because
in diagrams where the two photons couple to different constituents (4-point
contributions) the exchanged gluon goes on-shell within the range of
$x$-integration.
Such a situation occurs in the pure quark HSP as well. The usual treatment of
the
propagator poles
$
(g^2 \pm i \epsilon)^{-1} = {\cal P} (g^{-2}) \mp i \pi \delta
( g^2 ) \;
$
then leads to an imaginary part for the corresponding amplitudes. Together with
non-vanishing helicity-flip amplitudes (generated by the vector diquarks) this
gives
rise to a transverse polarization.

In Fig.\ref{compton} the diquark model result is confronted with experiment.
Also plotted is a prediction of the pure quark HSP for the
Chernyak-Ogloblin-Zhitnitsky
DA~\cite{Kron91}.
%

\subsection{$\gamma \gamma \longrightarrow \hbox{p}
\bar{\hbox{p}}$~\cite{Kro93a}}
\vspace{-0.35cm}
Simultaneously with Compton scattering also the crossed process $\gamma \gamma
\longrightarrow \hbox{p} \bar{\hbox{p}}$ has been considered. Unlike Compton
scattering there
occur no propagator singularities since Mandelstam $s$ and $t$ are now
interchanged. Shown in
Fig.\ref{ggppbar} is the integrated cross-section ($\vert \cos \theta \vert
\leq 0.6$) for
two-photon annihilation into p $\bar{\hbox{p}}$ as a function of $\sqrt{s}$.
For $\sqrt{s}
\ageq$  2.5 GeV good agreement with the recent CLEO data~\cite{Art94} can be
observed. The pure
quark model results of Farrar et al.~\cite{Fa85} lie down by about a factor of
8. Similar to the
magnetic form factor of the proton in the timelike region radiative (Sudakov)
corrections have
also been found to improve the situation in two-photon annihilation to a
certain
extend~\cite{Hy93}.
\begin{figure}
\vspace{7.5 truecm}
\fcaption{The integrated cross-section ($\vert \cos (\theta ) \vert \le 0.6$)
for $\gamma
\gamma \longrightarrow p \bar{p}$ as a function of $\sqrt{s}$ as resulting from
the
diquark model. Data are taken from Artuso et al.~\protect{\cite{Art94}}.
\label{ggppbar}}
\vspace{-0.5 truecm}
\end{figure}

\subsection{Photoproduction of Kaons~\cite{Schue92}}
\vspace{-0.35cm}

Photoproduction of mesons
$ \gamma p \rightarrow M B $
represents a large class of photon-induced reactions for which data at a few
GeV of
momentum transfer already exist~\cite{Shu79,And76}. Further data at even larger
$s$, $t$, and $u$
can be expected within the not-too-far future from the DESY e-p collider HERA.
{}From the
theoretical point of view, photoproduction means one step further in complexity
as compared to
Compton scattering. Within the diquark model the number of Feynman diagrams
contributing is in
general nearly three times larger (158) than for $ \gamma p \rightarrow \gamma
p$ (64). However,
there are fortunately a few photoproduction reactions for which matters become
somewhat simpler -
namely those where the baryon B in the final state is a $\Lambda$. Since the p
and the $\Lambda$
have in common only the S(ud) diquark such processes proceed solely through the
S diquark. For an
analogous reason only V diquarks are involved if the baryon becomes a
$\Sigma^0$. Since  $\gamma
p \rightarrow K^+ \Lambda$ only proceeds via S diquarks, scattering amplitudes
and hence
observables which require a flip of the baryonic helicity (like the proton
asymmetry or the Lambda
polarization) become zero.
Photoproduction of pseudoscalar mesons in general can be  described by 4
independent
helicity amplitudes
\be
   N = M_{-{1 \over 2}, 1 {1 \over 2}} \; , \qquad
   S_1 = M_{-{1 \over 2}, 1 -{1 \over 2}} \; , \quad
   D = M_{{1 \over 2}, 1 -{1 \over2}} \; ,   \qquad
   S_2 = M_{{1 \over 2}, 1 {1 \over2}} \; .
\ee
Since only scalar diquarks take part in $\gamma \hbox{p} \longrightarrow
\hbox{K}^+ \Lambda$
two of the 4 helicity amplitudes vanish, $ N = D = 0$. The remaining two
amplitudes, $S_1$ and
$S_2$, scale asymptotically like $s^{-5/2}$ and hence for large $s$ the
differential cross section
should decrease $\propto s^{-7}$.

The hard scattering amplitude $\widehat{T}$ for $\gamma p \rightarrow K^+
\Lambda$ is given by
all tree graphs for  $\gamma u S(ud) \longrightarrow u \bar{s} s S(ud)$,
altogether 63. As in
Compton scattering one encounters propagator singularities which give rise to
an imaginary part
in the helicity amplitudes. A new feature, however, showing up in
photoproduction is the
occurrence of triple gluon vertices.

\begin{figure}
\vspace{10.9 truecm}
\vspace{-2.2 truecm}
\fcaption{The photoproduction cross section vs. $\cos \theta$ for $\gamma p
           \rightarrow K^+ \Lambda$.
           The solid line represents the diquark-model result for the p and
$\Lambda$ DA of
           Eq.(\protect{\ref{DAp}}) and the $K^+$ DA of
Eq.(\protect{\ref{DAK}}). The dashed line represents a
           result obtained within the pure quark HSP~\protect{\cite{Fa91}}.
Data are
taken from
Ref.[25]
\label{photoprod}}
\end{figure}

With the $\Lambda$ DA given in Eq.(\ref{DAp}) (note, that this DA already
contains a flavour
dependence due to the constituent masses $m_q$ and $m_D$), an analogous form
for the K$^+$ DA
\vspace{-0.1 truecm}
\be
\phi^{K^+} (x_1) \propto x_1 (1-x_1) \exp
\left[ - b^2 \left( {m_u^2 / x_1} + {m_s^2 / (1-x_1)} \right) \right]
\; , \label{DAK}
\ee
\vspace{-0.1 truecm}
and $f_{K}$ fixed according to the experimental relation between $K$ and $\pi$
decay
constants, $f_{K} = 1.2 f_{\pi}$,  reasonable agreement with the data can be
achieved already
without introducing additional parameters (cf. Fig.\ref{photoprod}). The
predictions of the
pure quark HSP, also plotted in the figure, have been obtained with very
asymmetric DAs for
all hadrons involved. With the double humped Kaon DA used by these authors the
diquark-model
result can even be improved.

\vspace{-0.5 truecm}
\section{Outlook and Conclusions}
\vspace{-0.3 truecm}
For a phenomenological model, like the diquark model, it is certainly necessary
to  calculate as
many reactions as possible in a consistent way to check its usefulness.  For
the simplest
electron and photon-induced reactions (form factors, Compton scattering,
electroproduction of
Kaons, $\dots$) it already seems to work reasonably well. Presently, the
photoproduction of
arbitrary mesons, which requires the calculation of Feynman diagrams not only
for scalar, but
also for vector diquarks, is under investigation.  Unfortunately, exclusive
data in the few-GeV
region are often of poor quality and are mostly restricted to spin-averaged
quantities. Spin
observables would set, of course, the most stringent constraints. However, with
the increasing
number of reactions considered and with more and better exclusive data from new
facilities like
CEBAF it will undoubtedly be possible to clarify the role which diquarks play
in exclusive
scattering.

\section{Acknowledgement}
\vspace{-0.2 truecm}
Most of the results presented in this contribution are the outcome of a close
collaboration with R.Jakob, P.Kroll, M.Sch\"urmann (University of Wuppertal,
FRG) and
K.Passek (Rudjer Bo\v{s}kovi\`{c} Institute Zagreb, Croatia). I would like to
thank them
for many fruitful discussions and suggestions.

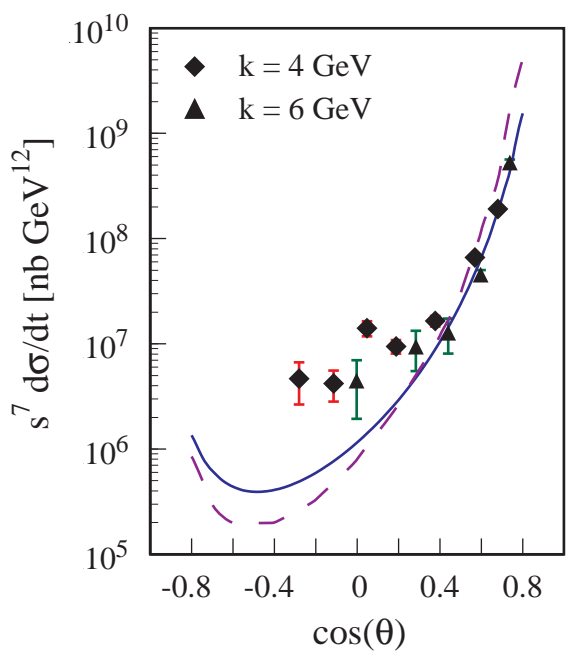
\end{document}